\documentclass[%
superscriptaddress,
preprintnumbers,
nofootinbib,
 amsmath,amssymb,
 aps,
prd,
]{revtex4-2}

\usepackage{graphicx}
\usepackage{subcaption}
\usepackage[justification=raggedright]{caption}
\usepackage{dcolumn}
\usepackage{bm}
\usepackage{hyperref}
\hypersetup{
  colorlinks   = true, 
  urlcolor     = blue, 
  linkcolor    = blue, 
  citecolor   = red 
}

\usepackage{slashed}
\usepackage{dsfont}

\usepackage{parskip}

\def\be{\begin{equation}}
\def\ee{\end{equation}}
\def\bea{\begin{eqnarray}}
\def\eea{\end{eqnarray}}
\def\bear{\begin{array}}
\def\ear{\end{array}}
\def\bfig{\begin{figure}}
\def\efig{\end{figure}}
\def\bcen{\begin{center}}
\def\ecen{\end{center}}
\def\bi{\begin{itemize}}
\def\ei{\end{itemize}}

\usepackage{bm}
\usepackage{xcolor}
\usepackage{multirow}
\usepackage{slashed}
\usepackage{physics}
\usepackage{makecell}

\newcommand{\no}{\nonumber}

\newcommand{\mL}{\mathcal{L}}
\newcommand{\mA}{\mathcal{A}}
\newcommand{\mB}{\mathcal{B}}

\newcommand{\mO}{\mathcal{O}}

\newcommand{\e}{\epsilon}

\newcommand{\cost}{\cos\theta}

\begin{document}

\title{The $\Lambda_c\to \Lambda\, \ell^+\nu_\ell$ weak decay
including new physics}

\author{Fernando Alvarado}
\email{Fernando.Alvarado@ific.uv.es}
\affiliation{Instituto de F\'isica Corpuscular (IFIC) and Departamento de F\'\i sica Te\'orica, \\ Universidad de Valencia and Consejo Superior de Investigaciones Cient\'{i}ficas (CSIC) 
\\ E-46980 Paterna, Valencia, Spain}
\author{Luis Alvarez-Ruso}%
\affiliation{Instituto de F\'isica Corpuscular (IFIC), \\  Consejo Superior de Investigaciones Cient\'{i}ficas (CSIC) and Universidad de Valencia \\ E-46980 Paterna, Valencia, Spain}
\author{Eliecer Hernández}
\affiliation{Departamento de F\'isica Fundamental and IUFFyM,
Universidad de Salamanca, E-37008 Salamanca, Spain}
\author{Juan Nieves}
\affiliation{Instituto de F\'isica Corpuscular (IFIC), \\  Consejo Superior de Investigaciones Cient\'{i}ficas (CSIC) and Universidad de Valencia \\ E-46980 Paterna, Valencia, Spain}
\author{Neus Penalva}
\affiliation{Institut de Ciencies del Cosmos (ICC), Facultat de F\'isica \\Universitat de Barcelona, E-08028 Barcelona, Spain}

\date{\today}

\begin{abstract}
We investigate the $\Lambda_c \to \Lambda \ell^{+} \nu_\ell$ decay with a focus on potential new physics (NP) effects in the $\ell = \mu$ channel. We employ an effective Hamiltonian  within the framework of the Standard Model Effective Field Theory (SMEFT) to consider generalized dimension-6 semileptonic $c\to s$ operators of scalar, pseudoscalar, vector, axial-vector and tensor types. We rely on Lattice QCD (LQCD) for the hadronic transition form factors, using heavy quark spin symmetry (HQSS) to determine those that have not yet been obtained on the lattice. Uncertainties due to the truncation of the NP Hamiltonian and different implementations of HQSS are taken into account. As a result, we unravel the NP discovery potential of the $\Lambda_c\to \Lambda$ semileptonic decay in different observables.  Our findings indicate high sensitivity to NP in lepton flavour universality ratios, probing multi-TeV scales in some cases. On the theoretical side, we identify LQCD uncertainties in axial and vector form factors as critical for improving NP sensitivity, alongside better SMEFT uncertainty estimations.
\end{abstract}

\maketitle

\section{\label{sec:intro} Introduction}

Hadron weak decays probe the flavour structure of matter, where unknown Beyond Standard Model (BSM) effects would manifest as differences between the SM prediction and precise measurements. The reader can consult Refs.~\cite{Goudzovski:2022scl,Artuso:2022ouk,Altmannshofer:2022hfs} for comprehensive reviews in both light and heavy quark sectors. In recent years, experimental hints to lepton flavour universality  (LFU) violation observed in $b \to c \ell \bar{\nu}_\ell$ with different hadrons in the initial and final states have received considerable attention. Theoretical studies and experimental searches have also been extended to the weak decays of $c$ and $s$ quarks. Although, for some quantities, the deviations from the SM have become smaller (a recent update can be found in Ref.~\cite{Iguro:2024hyk}), the theoretical strategies developed to identify and predict promising observables are valuable tools to keep putting the SM under scrutiny. 

Baryon decays are complementary to meson ones, providing a different, richer, sensitivity to new physics (NP), particularly when the study of decays rates is generalized to consider angular distributions and asymmetries.  
The $\Lambda_b\to \Lambda_c\, \ell^- \bar{\nu}_\ell$ decay has been extensively investigated in the SM and beyond~\cite{Dutta:2015ueb,Li:2016pdv,DiSalvo:2016rfi,Datta:2017aue,DiSalvo:2018ngq,Ray:2018hrx,Bernlochner:2018bfn,Hu:2018veh,Zhu:2018zxb,Ferrillo:2019owd,Boer:2019zmp,Penalva:2019rgt,Azizi:2019tcn,Mu:2019bin,Murgui:2019czp,Hu:2020axt,Penalva:2020xup,Penalva:2021gef, Penalva:2021wye, Penalva:2022vxy,Becirevic:2022bev,DiSalvo:2022pic,Fedele:2022iib,Geng:2022hmf,Karmakar:2023rdt,Nandi:2024aia,DiSalvo:2024amu}.
Some  studies considered $\Lambda_b\to p\, \ell^- \bar{\nu}_\ell$ as well~\cite{Dutta:2015ueb,Ray:2018hrx,Khan:2022ylv}. 
In most cases, lattice QCD (LQCD) input has been used but quark model~\cite{Gutsche:2015mxa,Faustov:2016pal,Li:2021qod,Zhang:2022bvl,Becirevic:2020nmb}, heavy quark effective theory~\cite{Bernlochner:2023jkp} and sum rule~\cite{Zhao:2020mod,Duan:2022uzm,Miao:2022bga,Aliev:2023tpk} calculations have also been performed. The $\Lambda_b$ inclusive semileptonic decays were examined in Ref.~\cite{Colangelo:2020vhu}. The transitions from $\Lambda_b$ to $\Lambda_c^*$ excited states  were considered in Refs.~\cite{Gutsche:2018nks,Nieves:2019kdh,Papucci:2021pmj,Meinel:2021rbm,Du:2022ipt,Li:2022hcn,DiRisi:2023npw}.
The $\Lambda_c\to \Lambda\, \ell^+\nu_\ell$ decay is also discussed in Refs.~\cite{Li:2021qod,Becirevic:2022bev,Bolognani:2024cmr}. In the light-quark sector, the NP discovery potential of $\Lambda$, $\Sigma$ and $\Xi$ semileptonic decays has been explored in Ref.~\cite{Chang:2014iba}. From the perspective of BSM searches, rare $\Sigma \to p \ell \ell$ decays are also interesting~\cite{Geng:2021fog,Roy:2024hqg}. Asymmetries in the angular distributions of $\Sigma^0 \to \Lambda \gamma$ compared to the corresponding ones with antiparticles would give access to new sources of CP violation~\cite{Nair:2018mwa}.

New experimental results are the driving force for these theory developments. The study of hadron decays, including hyperon ones, proceeds at BESIII, Belle II and LHCb. The  experimental program of LHCb and Belle II encompasses a variety of $b \to s$ and $b \to c$ processes. BESIII, operating in a regime where $\Lambda_c^+ \bar{\Lambda}_c^-$ pairs are produced in the exclusive two-body channel with no additional particles, has recently performed updated measurements of $\Lambda_c\to\Lambda \, \ell^+ \,\nu_\ell$ decays with $\ell = e,\,\mu$~\cite{BESIII:2022ysa,BESIII:2023jxv}. 

We study the impact of NP contributions to the $\Lambda_c\to\Lambda \, \ell^+ \,\nu_\ell$ decay, relying on the general framework of Ref.~\cite{Penalva:2020xup} for general $q\to q'$ charged-current transitions involving left-handed neutrinos. We consider scalar, pseudo-scalar and right-handed  tensor NP interactions, as well as left- and right-handed corrections to the SM, with interaction strengths given by Wilson coefficients.  
The effective Hamiltonian is the most general dimension-six low-energy realization of BSM theories and, in particular, of the Standard Model Effective Field Theory (SMEFT)~\cite{Buchmuller:1982ye,Grzadkowski:2010es}.

The QCD input can be parametrized in terms of 
hadronic structure functions built from matrix elements of the operators involved. These  structure functions are written in terms of transition form factors, which depend on the masses of the initial and final particles and on the invariant mass of the outgoing lepton pair. These form factors are directly taken from lattice QCD determinations, when possible. Otherwise we resort to approximations to express the unknown form factors in terms of known ones. 

In line with the standard hypothesis of flavour hierarchy of NP, we assume that $\ell = e$ is well described by the SM and focus on the $\ell = \mu$ channel. The fact that (pseudo)scalar and tensor NP contributions are $\mO(m_\ell)$ and therefore suppressed in the $e$ channel further reinforces the assumption.

In Sec.~\ref{sec:H} we first present the effective Hamiltonian, define the observables under consideration, the QCD input, and present the NP contributions and their constrains. The sensitivity to BSM effects is explored in Sec.~\ref{sec:Ratios} for LFU ratios and in Ref.~\ref{sec:Cs} for the coefficients in the angular decomposition of the decay rate. The conclusions of our study are summarised in Sec.~\ref{sec:conclusion}.

\section{\label{sec:H} The $\Lambda_c\to \Lambda\, \ell^+\nu_\ell$ weak decay including new physics}

\subsection{Effective Hamiltonian}

To investigate the $\Lambda_c \to \Lambda$ semileptonic decay we rely on the most general low-energy dimension-six effective Hamiltonian for the $c \to s \, \ell^+ \,\nu_\ell$ transition, considering only left-handed neutrinos and lepton-flavour conserving transitions (see for instance Ref.~\cite{Colangelo:2021dnv}) 
\begin{eqnarray}\label{eq:H}
    H_{\rm eff}&=&\frac{4 G_F}{\sqrt2} V_{c s}  \Big[
(1+\epsilon^\ell_L)\,(\bar s\gamma_\alpha
P_L c)\ (\bar\nu_\ell P_R\gamma^\alpha \ell)+
\epsilon^\ell_R\,(\bar s\gamma_\alpha
P_R c)\ (\bar\nu_\ell P_R\gamma^\alpha \ell)\nonumber\\
&&\hspace{2cm}+\frac{1}2\,\epsilon^\ell_S
(\bar s c)\ (\bar\nu_\ell P_R \ell)+\frac{1}2\,\epsilon^\ell_P
(\bar s \gamma_5 c)\ (\bar\nu_\ell P_R \ell)+\epsilon^\ell_T
(\bar s \sigma_{\alpha\beta}P_R c)\ (\bar\nu_\ell \sigma^{\alpha\beta}P_R \ell)\Big] \,, 
\end{eqnarray}
with $P_{L,R}=(1 \mp \gamma_5)/2$ denoting the projectors onto left and right
chiralities, respectively. $G_F$ is the Fermi decay constant and $V_{c s}$ the pertinent 
Cabibbo-Kobayashi-Maskawa (CKM) matrix element.  In the domain of applicability of SMEFT, where the NP scale $\Lambda$ is much higher than the electroweak one, the strength of NP terms is determined by flavour ($\ell$) dependent Wilson coefficients $\epsilon^\ell_{X}$, which are suppressed by $v^2/\Lambda^2$, where $v$ is the Higgs vacuum expectation value~\cite{Buchmuller:1982ye}. Although these Wilson coefficients are in general complex, we adopt the CP-conserving limit and treat them as real in the following. Sub-index $X=L$, $R$, $S$, $P$, $T$ denotes the left-handed, right-handed, scalar, pseudoscalar and tensor structure of the corresponding $c\to s$ transition operators\footnote{Notice that the Wilson coefficient for the term proportional to the left-handed quark current is denoted $\e^\ell_L$ differing with the notation of Ref.~\cite{Colangelo:2021dnv} where it is represented as $\e^\ell_V$.}. 
Note that, for left-handed neutrinos, a tensor operator with quark chirality $P_L$ instead of $P_R$ vanishes identically,  as can be shown with the help of Dirac algebra.

\par 

\subsection{Observables\label{sec:Obs}}

For the $\Lambda_c^+(p) \rightarrow \Lambda(p') \, \ell^+(k') \, \nu_l(k)$ decay, the double-differential branching ratio $\mB \equiv \Gamma / \Gamma_{\text{tot}}$, 
in terms of $q^2 = (k + k')^2$ and the angle $\theta$ between the $\Lambda$ and $\ell^+$ three-momenta in the leptonic rest frame, where 
$q^\mu = (\sqrt{q^2}, \vec{0})$,
can be cast as~\cite{Becirevic:2020rzi,Colangelo:2021dnv,Penalva:2020xup}
\begin{equation}\label{eq:ci}
\frac{{\rm d}^2\mB}{{\rm d}q^2 {\rm d} \cos\theta}= |V_{cs}|^2 \left[ c^\ell_0(q^2) +c^\ell_1(q^2) \cos\theta + c^\ell_2(q^2) \cos^2\theta \right]  \,,
\end{equation}
where $|V_{cs}|^2$ has been factorised for convenience. 
The full $\Lambda_c$ width $\Gamma_\text{tot}=(3.252\pm 0.050 )\times 10^{-12}$~GeV 
has been obtained from its mean life $\tau=(202.4\pm 3.1)\times 10^{-15}$~s~\cite{Zyla:2020zbs}. The functions  $c_i (q^2)$ entirely characterise the decay and can be extracted from experimentally measurable quantities. For example, as shown in Eqs.~\eqref{eq:diffwidth}-\eqref{eq:convexity}, they can be obtained from the single-differential decay width ${\rm d}\Gamma / {\rm d}q^2$, the forward-backward asymmetry, $\mA_{\rm{FB}}$, and the convexity, $\mA_{\pi/3}$. Our analysis of NP contributions is performed in terms of $c^\ell_{0,1,2}(q^2)$. We take advantage of the general framework developed in Refs.~\cite{Penalva:2020xup, Penalva:2021wye} for the study of hadron semileptonic decays in the presence of NP. Appendix~\ref{sec:Eliecer} discloses how the formalism derived in these references is adapted to the present case.

In order to reduce the theoretical and experimental uncertainties, it is useful to consider LFU ratios. In our case 
\begin{equation}
\label{eq:LFUrat}
    R_i\equiv\frac{c_i(m_\mu,q^2,\e^\mu)}{c_i(m_e,q^2,\e^e)}=\frac{c_i^\mu(q^2,\e^\mu)}{c_i^{e\rm{SM}}(q^2)}+\mO(\e^e)\underbrace{\simeq}_{\text{if }\e^e\ll\e^\mu}1+\mO(m_\mu^2/x^2)+\mO(\e^\mu) \ ,
\end{equation}
with $x^2=q^2,m_\Lambda^2,m_{\Lambda_c}^2$. The common assumption $\e^e\ll\e^\mu$ arises, for example, in many NP models which explain the fermion mass hierarchy~\cite{Barbieri:2011ci,Alonso:2016oyd}\footnote{In the context of dimension-six SMEFT, $\epsilon_R$ has been shown to be flavour independent~\cite{Cirigliano:2009wk,Alonso:2014csa,Cata:2015lta}. Therefore, within our assumption of flavour hierarchy, a non zero $\epsilon^\mu_R$ would be in conflict with SMEFT or come from a higher-order operator. We do not impose the flavour universality assigned to $\epsilon_R$ in the context of SMEFT. We refrain to do so to be more general, since $\epsilon_R$ could be flavour dependent in a nonlinear realization of the electroweak symmetry breaking~\cite{Cata:2015lta}.}. 
In these ratios, not only $V_{\rm CKM}$ cancels out  but also an important advantage resides in the fact that in the SM the deviation from unity enters at $\mO(m_\mu^2/x^2)$  and so does the SM uncertainty. 
By using ratios  a significant improvement in the sensitivity to NP is hence achieved, as compared to $c_i$ values themselves, where the SM uncertainties enter without any suppression.

\subsection{QCD input\label{sec:QCDinput}}

The functions $c_{0,1,2}(q^2)$ depend on the NP Wilson coefficients introduced above and on purely hadronic structure functions which are given in terms of transition form factors (see Appendix E of Ref.~\cite{Penalva:2020xup}). For a $J^P = 1/2^+\to 1/2^+$ transition, like the $\Lambda_c \to \Lambda$ one, there are 12 real $q^2$ dependent form factors: scalar and pseudoscalar $F_{S,P}$, vector $F_{1,2,3}$, axial-vector $G_{1,2,3}$ and tensor $T_{1,2,3,4}$. 
Adopting the notation of Ref.~\cite{Penalva:2020xup} and references therein
\begin{eqnarray}
 \langle\Lambda (p')| \bar s \left(1-\gamma_5\right) c |
 \Lambda_c (p)\rangle &=& 
 \bar{u}_{\Lambda} (p^{\prime})\left(F_S-\gamma_5 F_P\right) 
 u_{\Lambda_c}(p), \nonumber \\
\langle\Lambda (p^\prime) | \bar s \gamma^\alpha\left(1-\gamma_5\right) c |
 \Lambda_c (p)\rangle &=& 
\bar{u}_{\Lambda}(p^\prime)\left\{ \gamma^\alpha 
\left(F_1-\gamma_5G_1\right)  + \frac{p^{\alpha}}{M} 
\left(F_2-\gamma_5G_2\right) +
 \frac{p^{\prime\alpha}}{m}\left(F_3-\gamma_5G_3\right) \right\} 
 u_{\Lambda_c}(p) \,, 
 \\
\langle\Lambda (p^\prime)| \bar s \sigma^{\alpha\beta} c | \Lambda_c (p)\rangle &=& 
 \bar{u}_{\Lambda}(p^{\prime})
 \left\{\frac{i}{M^2}
 \left(p^\alpha p^{\prime \beta}-p^\beta p^{\prime \alpha}\right) T_1
 + \frac{i}{M}\left(\gamma^\alpha p^\beta
  -\gamma^\beta p^\alpha  \right) T_2 \right.
 \nonumber 
 \\ 
 &&\hspace{1.1cm}
 \left. 
 + \frac{i}{M}\left(\gamma^\alpha p^{\prime \beta}
-\gamma^\beta p^{\prime \alpha}\right) T_3
+ \sigma^{\alpha\beta} T_4 \right\} u_{\Lambda_c}(p) \,, \nonumber
\label{eq:FactoresForma}
\end{eqnarray}
where $M(m)$ denotes the $\Lambda_c$($\Lambda$) mass. The vector and axial form factors have been calculated within LQCD in Ref.~\cite{Meinel:2016dqj} using gauge field configurations generated by the RBC and UKQCD collaborations with $2+1$ flavours of dynamical domain-wall fermions\footnote{A recent extraction of these form factors in the framework of QCD sum rules is also of interest~\cite{Zhang:2023nxl}.}. Two different lattice spacings are analysed and the physical pion mass is reached for one ensemble. The translation of the form factors from the helicity basis adopted in Ref.~\cite{Meinel:2016dqj} to the one adopted here is given by Eq.~(E1) of  Ref.~\cite{Penalva:2020xup} once the baryon masses are changed accordingly. Using these form factors, in agreement with Ref.~\cite{Meinel:2016dqj} and propagating the statistical uncertainties according to the correlation matrices provided in Ref.~\cite{Meinel:2016dqj} we obtain: 
\begin{equation}
    \frac{\Gamma(\Lambda_c\to\Lambda e^+ \nu_e)}{\abs{V_{cs}}^2}=0.2008(71)^{\rm LQCD}_{\rm{stat}}\ \rm{ps}^{-1},\quad   \frac{\Gamma(\Lambda_c\to\Lambda \mu^+ \nu_\mu)}{\abs{V_{cs}}^2}=0.1945(69)^{\rm LQCD}_{\rm{stat}}\ \rm{ps}^{-1},\ 
\end{equation}
yielding the LFU ratio 
\begin{equation}\label{eq:RGamma}
    R_{\mathrm{SM}}=\frac{\Gamma(\Lambda_c\to\Lambda \mu^+ \nu_\mu)}{\Gamma(\Lambda_c\to\Lambda e^+ \nu_e)}=0.96884(61)^{\rm LQCD}_{\rm{stat}}\ .
\end{equation}
We notice the difference between this SM ratio and the value of $0.974(1)$ reported in Ref.~\cite{Becirevic:2022bev} although both are consistent with the experimental result of  $R^{\rm exp}=0.98(5)_{\rm stat}(3)_{\rm syst}$~\cite{BESIII:2023jxv} 
owing to its large error.  
 
Ideally one would also take the rest of form factors from LQCD, relying in this way on an accurate parametrisation of the QCD structure. However this input is not yet available.
  Therefore, we express the scalar, pseudoscalar and tensor form factors in terms of the vector and axial ones employing approximations based on the fact that the charm quark mass is considerably larger than the masses of the light ones: $m_c \gg \Lambda_{\rm{QCD}} \gg m_{s,u,d}$. 
There is some freedom in the way one exploits this condition. The resulting differences are used to estimate the systematic uncertainty of the approximations. As shown in Sec.~(2.10) of Ref.~\cite{Manohar:2000dt}, HQSS for a heavy-light quark transition implies that at leading order in $1/m_c$ all possible form factors can be given in terms of only two independent functions. For this purpose we choose either $F_{1,2}$ or $G_{1,2}$  ($F_3=G_3=0$ in the HQSS limit). We refer to these prescriptions as HQSSV and HQSSA respectively. A less stringent choice arises from the assumption that the $c$ quark is on-shell. Then, in the $m_c/M \to 1$ limit, one obtains a consistent set of equations which allow to relate all (pseudo)scalar and tensor form factors to $F_{1,2,3}$ and $G_{1,2,3}$. This solution, labelled OSHQ (for On-Shell Heavy Quark), is consistent with HQSS in the $F_3=G_3=0$ limit. The three prescriptions are listed below. Details of the derivations can be found in Appendix~\ref{sec:HQSS}. 
\begin{enumerate}
    \item OSHQ:
    \begin{eqnarray}\label{eq:ffLuis}
        F_S&= F_1+F_2+\omega F_3  \ ,\qquad\qquad & F_P= G_1-G_2-\omega G_3 \ ,\no\\
     T_1&=  -\frac{M}{m} \left(F_3+G_3\right) \ ,\qquad\qquad & T_2=F_1-G_1-G_3  \ ,\no\\
     T_3&= \frac{M}{m} G_3 \ ,\qquad\qquad\qquad\qquad & T_4= G_1-\left( \omega-1 \right)G_3 \ ,
    \end{eqnarray}
    where
    $\omega=(M^2+m^2-q^2)/(2 M m)$ is the scalar product of the four-velocities of the initial and final hadrons.
    \item HQSSV:
    \begin{equation}\label{eq:ffV}
        F_S= F_1+F_2 \ ,\quad F_P= F_1 \ ,\quad T_1= 0 \ ,\quad T_2 =-F_2 \ ,\quad  T_3= 0 \ ,\quad  T_4= F_1+F_2 \ .
    \end{equation}

    \item HQSSA:
    \begin{equation}\label{eq:ffA}
        F_S= G_1 \ ,\quad F_P=G_1-G_2 \ ,\quad T_1= 0 \ ,\quad  T_2 = - G_2 \ ,\quad T_3= 0 \ ,\quad\ T_4= G_1 \ .
    \end{equation}
\end{enumerate}

\begin{figure}[h]
\begin{subfigure}[t]{0.32\textwidth}
    \centering\includegraphics[width=\textwidth]{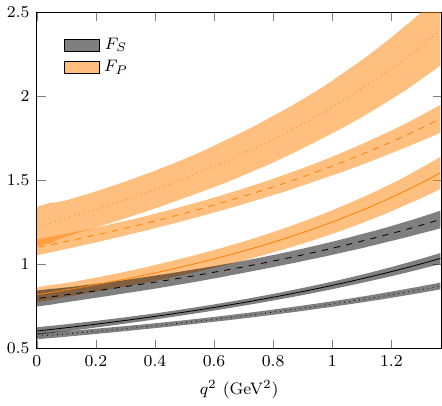}
    \caption{}
  \end{subfigure}\hfill
  \begin{subfigure}[t]{0.32\textwidth}
    \centering\includegraphics[width=\textwidth]{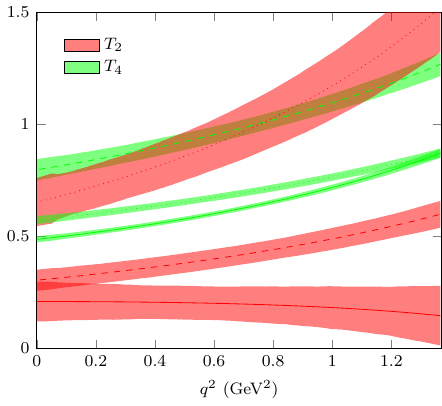}
   \caption{}
  \end{subfigure}\hfill
  \begin{subfigure}[t]{0.32\textwidth}
    \centering\includegraphics[width=\textwidth]{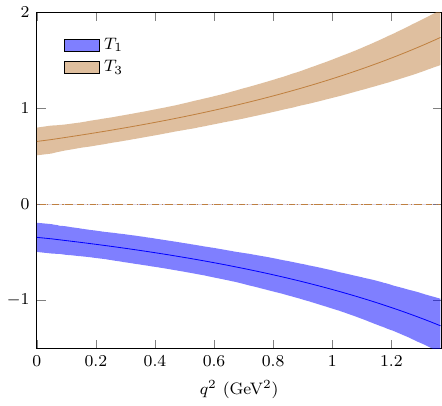}
    \caption{}
  \end{subfigure}
  \caption{Scalar, pseudoscalar and tensor form factors obtained from the vector and axial ones using different approximations. The continuous, dashed and dotted lines are obtained with OSHQ, HQSSV and HQSSA prescriptions, respectively.  
  Notice that $T_1=T_3=0$ within HQSSV and HQSSA.
  }
\label{fig:ff}
\end{figure}
Figure~\ref{fig:ff} displays $F_{P,S}$ and $T_{1-4}$ in terms of $F_{1-3}$ and $G_{1-3}$ from LQCD~\cite{Meinel:2016dqj} in the $q^2$ range of relevance for the $\Lambda_c$ decay using the three prescriptions. In the following, our theoretical error bands for each of these form factors cover the results of all three prescriptions including the LQCD errors given by the bands in Fig.~\ref{fig:ff}. 
This defines a conservative estimate of the uncertainty in the $S$, $P$  and $T$ form factors, derived using different prescriptions based on leading order HQSS.

\subsection{\label{sec:cix} New physics contributions}
In order to analyse individual NP contributions to $c^\ell_{0,1,2}$ we introduce 
\begin{equation}
  \label{eq:ciXdef}  
    c^\ell_i(q^2,\e^\ell_X) \equiv c^\ell_i(q^2,\epsilon^\ell_X,\e^\ell_{Y\neq X}=0)\,,\qquad  X,Y=L,R,S,P,T\,, \, i =0,1,2
\end{equation}
which can be cast as (lepton index and $q^2$ dependence are implicit)
\bea\label{eq:ciX}
    c_i(\epsilon_X)&=& c_i^\text{SM}+\epsilon_X c_i^{X\rm{l}}+\epsilon_X^2 c_i^{X\rm{q}}=c_i^\text{SM}+\epsilon_X\left( c_i^{X\rm{l}}+\epsilon_X c_i^{X\rm{q}} \right)\equiv c_i^\text{SM}+\epsilon_X c^X_i(\epsilon_X) \ .
\eea
The NP part has terms linear and quadratic in $\e_X$. The leading linear part $\e_X c_i^{X\rm{l}}$  arises from the interference between SM and NP terms in the decay amplitude squared.  On the other hand, one has to keep in mind that the sub-leading $\e^2_X c_i^{X\rm{q}}$ terms provide an incomplete account of the $\mO(\e^2)$ corrections, which would also receive contributions from the interference between the SM and terms from a higher order $(\e^2)$ SMEFT Hamiltonian, controlled by additional Wilson coefficients~\cite{Murphy:2020rsh}. 

\begin{figure}[h]
\begin{subfigure}[t]{0.32\textwidth}
    \centering\includegraphics[width=\textwidth]{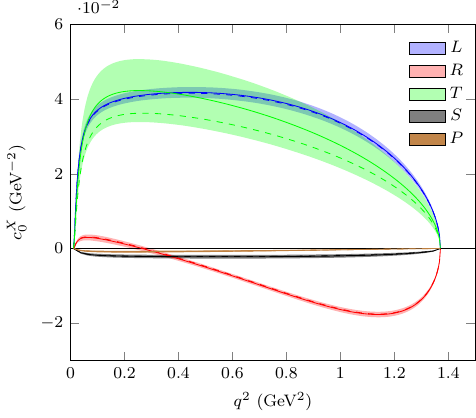}
    \caption{}
  \end{subfigure}\hfill
  \begin{subfigure}[t]{0.32\textwidth}
    \centering\includegraphics[width=\textwidth]{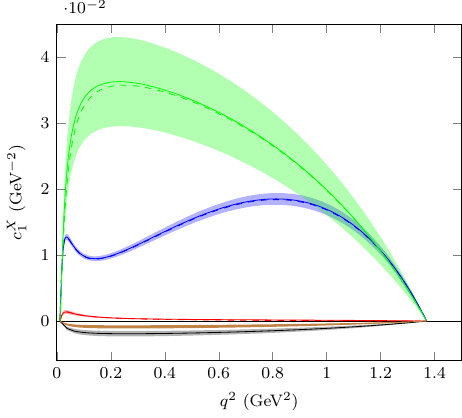}
   \caption{}
  \end{subfigure}\hfill
  \begin{subfigure}[t]{0.32\textwidth}
    \centering\includegraphics[width=\textwidth]{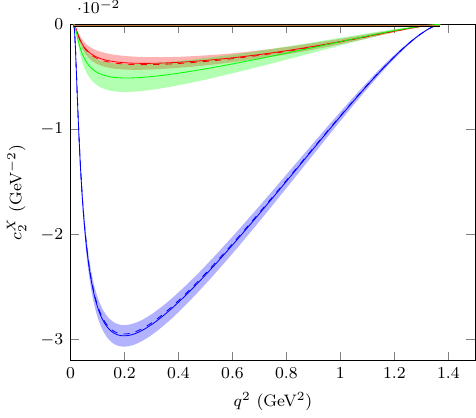}
    \caption{}
  \end{subfigure}
  \caption{$q^2$ dependence of the NP individual perturbations $c^X_i=c_i^{X\text{l}}+\epsilon_X c_i^{X\text{q}}$, Eq.~\eqref{eq:ciX} ($i=0,1,2$ from left to right). Each colour denotes a type of interaction: $X = L$ (blue), $R$ (red), $S$ (black), $P$ (brown) and $T$ (green). Wilson coefficients are set to their central values in Ref.~\cite{Becirevic:2020rzi} (see text).  The dashed lines originate from the linear terms $c_i^{X\rm{l}}$. The bands represent the QCD uncertainty from the transition form factors.}
\label{fig:ciX}
\end{figure}
The structure of the NP perturbation of the SM values is therefore encoded in $c^X_i(\epsilon_X)$, which are depicted in Fig.~\ref{fig:ciX} for $\ell = \mu$. Propagated QCD uncertainties, from LQCD and HQSS prescriptions discussed in Sec.~\ref{sec:QCDinput}, give rise to the bands. The dashed lines denote the linear $c_i^{X\rm{l}}$ terms that contribute at $\mO(1/\Lambda^2)$. Although for $X=S, P, T$ these interference terms are suppressed by the muon mass, the tensor contribution turns out to be comparable or larger than the $X=L, R$ ones for $c_0$ and $c_1$ observables. The origin of this behaviour can be traced back to larger numerical factors present in the hadron matrix elements of the tensor operator. To account for the residual $\e_X$ dependence of $c^X_i(\epsilon_X)$, we choose as indicative the central $\e_X$ values extracted in Ref.~\cite{Becirevic:2020rzi} from leptonic and semileptonic decays of pseudoscalar mesons. 
Under the assumption of real Wilson coefficients, the relations between the definitions in Ref.~\cite{Becirevic:2020rzi} ($g$) and the ones in the present work ($\e$) are  
\begin{equation}
    \e_{L,R}=\frac{g_V\mp g_A}{2}=g_{V_{L,R}},\quad \e_{S,P,T}=g_{S,P,T} \,.
\end{equation}
We therefore obtain the following ranges: $\e_L=(1.2\pm 2.0)\times 10^{-2}$, $\e_R=(-0.9\pm 2.0)\times 10^{-2}$, $\e_S=(-1\pm 2)\times 10^{-2}$, $\e_P=(0.7\pm 1.4)\times 10^{-3}$, $\e_T=(1.2\pm 1.8)\times 10^{-2}$.\footnote{These values differ from those in Ref.~\cite{Colangelo:2021dnv} although they are obtained from the same source.}

The smallness of the quadratic terms is apparent from the difference between the dashed and the central solid lines, which can hardly be discerned except for $c_{0,2}^T$. Indeed, $\epsilon_T c_0^{T\text{q}}$ is relatively large for a typical value of $\e_T=1.2\times 10^{-2}$. The tensor contribution to $c_2$ is purely quadratic ($c_2^{T\text{l}} = 0$)  and quite large. $c_0$ is the only observable where the quadratic, $c_{0}^{S\rm{q}}$ and $c_{0}^{P\rm{q}}$, scalar and pseudoscalar contributions  are exactly zero. 

The plots of Fig.~\ref{fig:ciX}  show the magnitude and shape of the individual contributions. They allow us to identify the NP perturbations that are in principle more easily revealed in an experiment. The sensitivity to these NP contributions is carefully studied in Sec.~\ref{sec:sens}. To assess the results of Fig.~\ref{fig:ciX} from this perspective it is useful to bear in mind that $c_i^\text{SM} = c_i^{L\rm{l}}/2$. For $c_0$ [Fig.~\ref{fig:ciX} (a)], tensor and left-handed contributions prevail, while the right-handed one is also sizeable and changes sign.  In $c_1$ [Fig.~\ref{fig:ciX} (b)] the tensor interaction is the largest, peaking in the $q^2$ region where the SM has a local minimum. Since the SM relative error is quite constant in $q^2$, a minimum in the SM magnitude corresponds to a minimum in the SM error (see the width of the $c_1^{L}$ band). Therefore one can expect a good sensitivity to tensor NP at low $q^2$ in $c_1$. Finally, in $c_2$ [Fig~\ref{fig:ciX} (c)] the left-handed term dominates, having a maximum in its absolute value at low $q^2$. The suppression of all contributions with respect to the left-handed one implies that the latter is the only potentially relevant NP mechanism in $c_2$. 
 It is also worth noting that the scalar and pseudoscalar contributions linear in $\e_{S,P}$ are small in $c_{0,1}$ and exactly zero in $c_2$.

\subsection{\label{sec:pheno}Phenomenological constraints}
\begin{figure}[h!]
\begin{subfigure}[t]{0.32\textwidth}
    \centering\includegraphics[width=\textwidth]{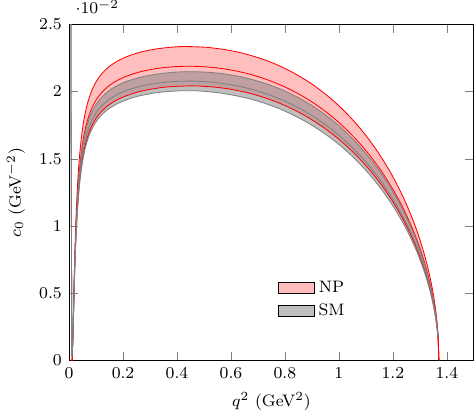}
    \caption{}
  \end{subfigure}\hfill
  \begin{subfigure}[t]{0.32\textwidth}
    \centering\includegraphics[width=\textwidth]{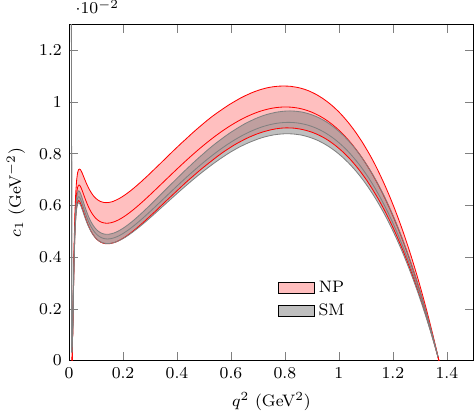}
   \caption{}
  \end{subfigure}\hfill
  \begin{subfigure}[t]{0.32\textwidth}
    \centering\includegraphics[width=\textwidth]{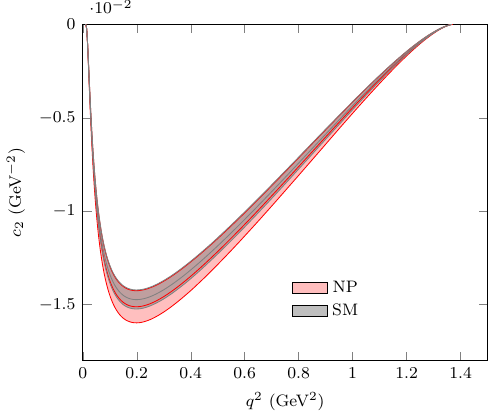}
    \caption{}
  \end{subfigure}
  \begin{subfigure}[t]{0.32\textwidth}
    \centering\includegraphics[width=\textwidth]{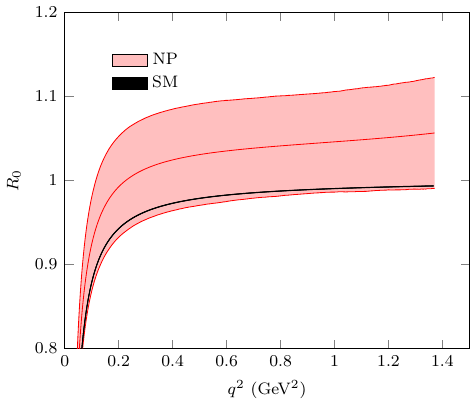}
    \caption{}
  \end{subfigure}\hfill
  \begin{subfigure}[t]{0.32\textwidth}
    \centering\includegraphics[width=\textwidth]{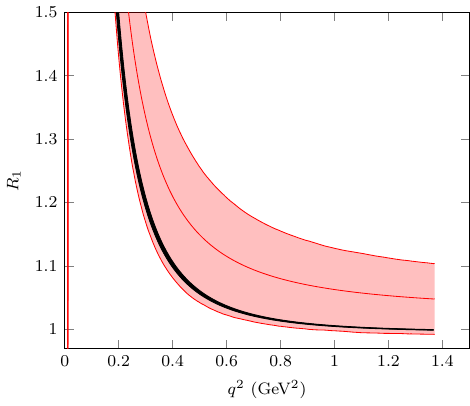}
    \caption{}
  \end{subfigure}\hfill
  \begin{subfigure}[t]{0.32\textwidth}
    \centering\includegraphics[width=\textwidth]{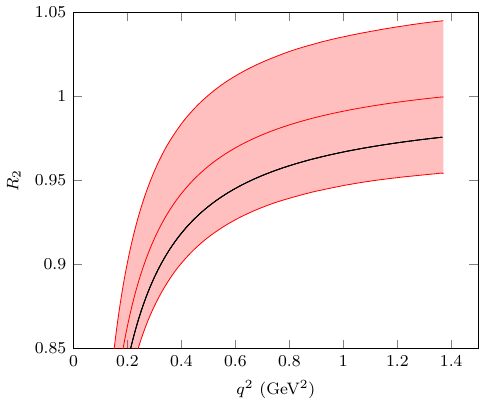}
    \caption{}
  \end{subfigure}
  \caption{$q^2$ dependence of $c^\mu_i$ and $\mu/e$ LFU ratios $R_i$ in the SM and in the presence of NP (with the empirical values obtained in Ref.~\cite{Becirevic:2020rzi}). The SM bands (grey) include errors in the LQCD vector form factors. The NP bands (red) arise from both the uncertainties in all form factors and in the Wilson coefficients.}
\label{fig:ciPenuelas}
\end{figure}

In the first row of Fig.~\ref{fig:ciPenuelas} functions $c_i(q^2)$ are plotted in the SM and also adding NP according to the aforementioned phenomenological constraints to the Wilson coefficients from Ref.~\cite{Becirevic:2020rzi} in the $\mu$ channel. We account simultaneously for all NP effects,  considering  linear terms as well as quadratic ones, which include interference $\mO(\e_X\e_Y$) contributions.
We have varied the $\e_X$ values in the given ranges, approximating their probability distributions by Gaussians and neglecting correlations. The LQCD errors of the vector and axial form factors have been propagated according to the correlation matrices given in Ref.~\cite{Meinel:2016dqj}. All three prescriptions of the HQSS form factors in $c_i^{S,P,T}$ are encompassed by a single band.\footnote{In this case the HQSSV choice practically covers the bands of the other two prescriptions.} Given that the Wilson coefficients under study are compatible with zero, the SM and NP bands overlap even in the absence of QCD uncertainties.  Nevertheless, one can already see where the possible NP contribution is more prominent. Figure~\ref{fig:ciPenuelas} (b) reflects that the SM local minimum is a good place to look for a NP 
(tensor) signal. Indeed, it is apparent that at this local minimum the SM and NP central values are more separated than anywhere else. 

\begin{figure}[h!]
    \centering
    \includegraphics[width=0.35\textwidth]{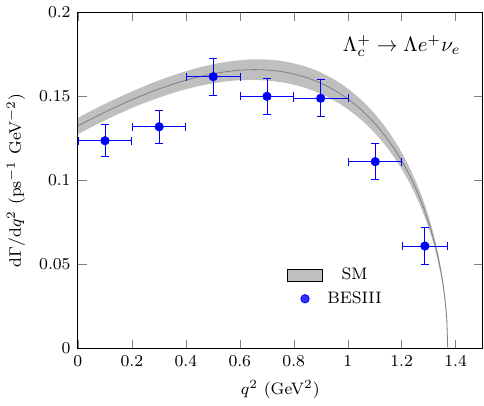}\hspace{1cm}
    \includegraphics[width=0.35\textwidth]{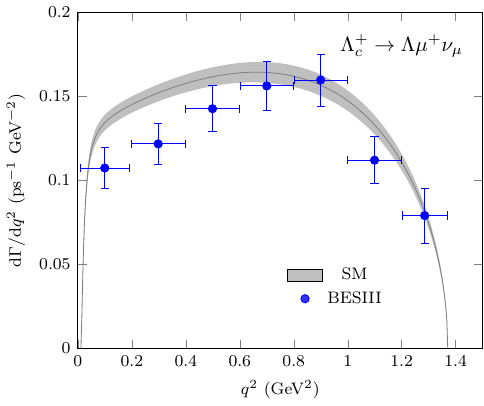}
    \caption{Differential decay width for 
    $\Lambda_c^+\to \Lambda e^+\nu_e$ (left) and $\Lambda_c^+\to \Lambda \mu^+\nu_\mu$ (right) decay modes.  The grey band is the SM prediction with 1$\sigma$ errors from the LQCD form factors (neglecting 1\% error in $V_{cs}$). The blue points are the experimental data from BESIII~\cite{BESIII:2022ysa,BESIII:2023jxv}.}
    \label{fig:dGamm}
\end{figure}

Under the assumption of the standard hierarchy, $\e_X^\mu\gg \e_X^e$, one would expect to observe deviations from the SM only in the muon channel. However, the comparison of the SM differential decay width, evaluated with LQCD input, to the BESIII data of Refs.~\cite{BESIII:2022ysa,BESIII:2023jxv} shown in Fig.~\ref{fig:dGamm} reveals tensions not only in the $\ell=\mu$ mode but also in the $\ell=e$ one.
Moreover, the SM fits performed in Refs.~\cite{BESIII:2022ysa} produce form factors that clearly differ from those of Ref.~\cite{Meinel:2016dqj}, as can be seen in Fig.~3 of Ref.~\cite{BESIII:2022ysa}. We have also obtained that this tension with experimental data cannot be accommodated by natural $\e^e_X \neq 0$.

Therefore, the extraction of allowed ranges on $\epsilon^\mu_X$ under the assumption of $\e^e\ll\e^\mu$ directly from the present data on $\Lambda_c\to\Lambda \mu^+ \nu_\mu$~\cite{BESIII:2023jxv} would be undermined by these discrepancies in the $\ell=e$ channel that cast doubts about the QCD input.
\par 
Turning now to the LFU violation ratios, Eq.~(\ref{eq:LFUrat}), setting $\e^e_X=0$ one gets
\begin{equation}\label{eq:Rdef}
    R_{i}(\e^\mu_X)=\frac{c_i(\Lambda_c\to\Lambda \mu^+ \nu_\mu,\e^\mu_X)}{c_{i}^{\rm SM}(\Lambda_c\to\Lambda e^+ \nu_e)} \ ,
\end{equation}
which depend on all $\e^\mu_X$, and implicitly on $q^2$. They are shown in the second row of Fig.~\ref{fig:ciPenuelas} with $\e^\mu_X$  values set to the range of the phenomenological constraints from Ref.~\cite{Becirevic:2020rzi}. As anticipated, the reduction of the SM uncertainty is evident: the SM error bands become negligible compared with the bands generated by the $\e^\mu_X$ ranges. This provides a much higher sensitivity to NP.

\section{\label{sec:sens}Sensitivity to new physics in the $\mu$ channel: LFU ratios}
\label{sec:Ratios}

We have posed ourselves the question of how large should a new physics contribution be in order to be experimentally identifiable, given the QCD uncertainties. We focus on the $\mu$ channel, under the assumption $\e^\mu\gg\e^e\simeq0$, and study ratios to reduce the uncertainties.   We have defined the signals to which we are sensitive as the ones that do not overlap with the SM prediction within $1\sigma$ errors. In other words, if a Wilson coefficient has a given true value, $\e_X$, we would be sensitive to it at a certain $q^2$ only if its magnitude is large enough to completely separate the $1\sigma$ NP band from the SM one. This condition has some degree of arbitrariness and would still be insufficient to claim a discovery in an experiment but serves the purpose of getting an idea of the NP discovery potential of each of the LFU ratios, $R_i$, as functions of $q^2$. We define  $\e_X^{\rm{sens}}$ as the minimum, in magnitude, $\e_X$ value which satisfies this condition. We evaluate them one by one, setting the rest of the Wilson coefficients to zero. The combined effect of more than one NP operator has been studied in the literature, e.g.  Refs.~\cite{Colangelo:2021dnv,Murgui:2019czp,Bolognani:2024cmr}. We leave the
simultaneous analysis for future work, once the form factors are determined more precisely. Recalling the notation of Eq.~\eqref{eq:Rdef} we define $\e_X^{\rm{sens}}$ as the one that satisfies
\begin{equation}\label{eq:sens}
    \abs{R_i(\e_X)-R_i^{\rm{SM}}}>\Delta R_i(\e_X)+\Delta R_i^{\rm{SM}}
\end{equation}
with a minimal $|\e_X|$. Here $\Delta R_i^{SM}$ and $\Delta R_i(\e_X)$ denote the $1\sigma$ errors of $R_i^{\rm{SM}}$ and $R_i(\e_X)$, which includes NP. On the one hand, the larger the $\e_X$, the more promising the observable would be to detect a given NP signal. On the other hand, the SMEFT calculation has a worse convergence 
for larger $\e_X$. We take this into account by introducing a SMEFT truncation error within $\Delta R_i(\e_X)$, which is defined below.

We explore the condition of Eq.~\eqref{eq:sens} for
positive and negative $\e_X$ separately. In general we do not obtain the same absolute value  in both cases due to quadratic corrections. 
Quadratic terms in SMEFT are briefly discussed in Ref.~\cite{Corbett:2021jox}, while the theoretical uncertainty of SMEFT is addressed in Sect.~2.7.6 of Ref.~\cite{Passarino:2016pzb}. According to this reference, to properly account for the truncation error one should calculate to one-loop order. It is also argued that an estimate can be given without such a computation, but including Wilson coefficients of the dimension-eight Lagrangian (Eq.~(2.79) therein). Such studies are beyond the scope of the present work. 
We content ourselves with a simpler approach, relying on a  common truncation error estimate given by the difference between linearization and quadratization of the matrix element squared, and define the following uncertainty for $c_i$, Eq.~(\ref{eq:ciX})
\begin{equation}\label{eq:err}
\Delta c_i^\text{SMEFT}(\e_X) =\e_X^2\max\{| c_i^{\rm{SM}}|,| c^{X\rm{l}}_i|,|c^{X\rm{q}}_i|\} \,.
\end{equation}
It estimates the size of the unaccounted $\mO(\e^2)$ contributions, only with information from $\mL_{\rm{SM}}$ and the NP dimension-six  Lagrangian $\mL_6$. To simplify the notation, index $\ell = \mu$ has been omitted throughout this section. Correspondingly, 
\be
\label{eq:errRatio}
\Delta R_i^\text{SMEFT}(\e_X)=\frac{\Delta c_i^\text{SMEFT}(\e_X)}{c_i^{e{\rm SM}}} \ .
\ee
 We consider this  a conservative estimation of the truncation error.
\par 
In order to obtain $\e_X^{\rm{sens}}$ from Eq.~\eqref{eq:sens} we sum in quadratures the error propagated from the LQCD vector and axial form factors, $\Delta R_i^{\rm{LQCD}}$, with the truncation one, yielding
\be
\label{eq:errQuad}
\left(\Delta R_i(\e_X)\right)^2=\left(\Delta R_i^{\rm{LQCD}}(\e_X)\right)^2+\left(\Delta R_i^\text{SMEFT}(\e_X)\right)^2 \ .
\ee
In addition, the HQSS uncertainty in $c_i^{S,P,T}$ is taken into account by covering all three versions of the form factors introduced in Sec.~\ref{sec:pheno}. In this case, this reduces to taking the least favourable choice, namely the HQSS prescription associated with the minimum perturbation to the SM, maximising therefore $\abs{\e_X^{\rm{sens}}}$.

First of all, we study the ratio of integrated widths, \textit{i.e.} we introduce NP contributions in the numerator of Eq.~(\ref{eq:RGamma}). In this observable,  $\abs{\e_L^{\rm{sens}}}=6.3\times 10^{-4}$, $\abs{\e_R^{\rm{sens}}}=2.1\times 10^{-3}$, $\abs{\e_T^{\rm{sens}}}=8.0\times 10^{-4}$, $\abs{\e_S^{\rm{sens}}}=8.9\times 10^{-3}$, but we are not sensitive to $\e_P$. This is a very good sensitivity, beyond the multi-TeV scale. Indeed, the $\e$ value corresponding approximately to a NP scale $\Lambda$ of 1 TeV is estimated by  a naive power counting (see Sec. 2.1 of Ref.~\cite{Passarino:2016pzb}), to be $\abs{\e}=v^2/\Lambda^2\simeq0.0625$, with $v\approx 250$~GeV.

Next, we analyse the LFU ratios defined in Eq.~\eqref{eq:Rdef}. The first thing to notice is that the SM $R_2$ is independent of the form factors. Therefore $\Delta R_2^{\rm SM}=0$. Applying the criterion of Eq.~\eqref{eq:sens}, we are sensitive to arbitrarily small NP $\e_{L,R,T}$, which are the ones that contribute to $c_2$. This has to be understood as a result of neglecting $\mO(\e^2)$ 
(and disregarding the experimental uncertainty). Nevertheless, this result clearly indicates that $R_2$ is a very sensitive observable to NP contributions. The drawback is that it might be difficult to measure, because $c_2$ is related to the angular asymmetry $\mA_{\pi/3}$, Eq.~(\ref{eq:convexity}).

\begin{figure}[h]
\begin{subfigure}[t]{0.40\textwidth}
    \includegraphics[width=\textwidth]{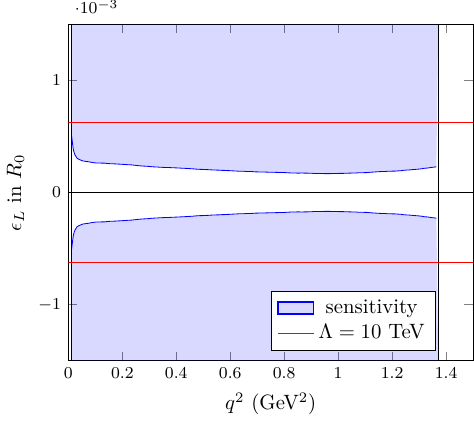}
    \caption{}
  \end{subfigure}\hspace{0.7cm}
  \begin{subfigure}[t]{0.40\textwidth}
    \includegraphics[width=\textwidth]{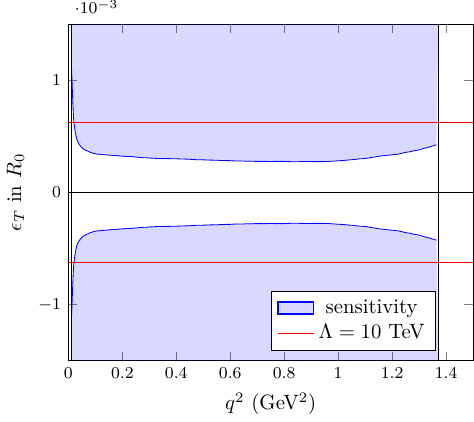}
    \caption{}
  \end{subfigure}\vfill
  \begin{subfigure}[t]{0.40\textwidth}
    \includegraphics[width=\textwidth]{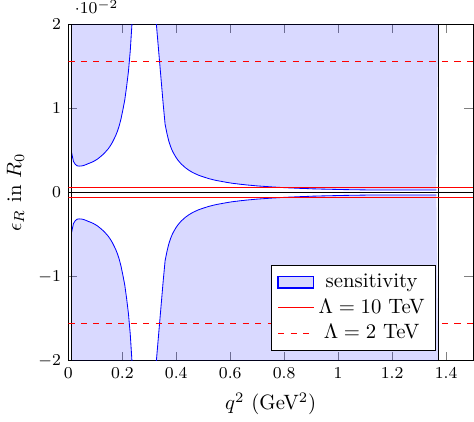}
    \caption{}
  \end{subfigure}\hspace{0.7cm}
  \begin{subfigure}[t]{0.40\textwidth}
\centering\includegraphics[width=\textwidth]{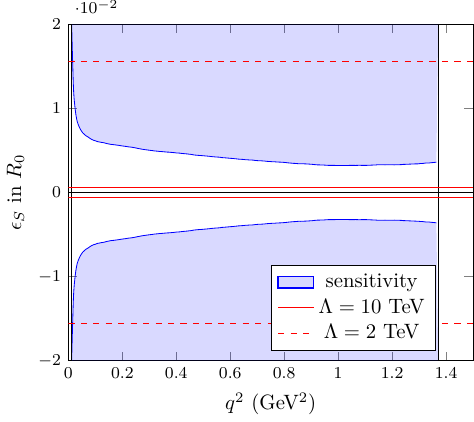}
\caption{}
  \end{subfigure}
\caption{The sensitivity regions to $\e_X$ in $R_0$ as a function of $q^2$ are depicted in light blue. The solid blue line represents $\e_X^{\rm{sens}}$ defined in Eq.~\eqref{eq:sens}. Currents of chirality $L$, $T$, $R$ and $S$ are studied in (a), (b), (c) and (d), respectively. Notice the different scale of the ordinate axis in upper and lower panels. The red solid (dashed) lines represent the Wilson coefficient value for a $\Lambda=10$ TeV ($\Lambda=2$ TeV) in the naive power counting. Black vertical lines delimit the phase space. $R_0$ is sensitive to Wilson coefficients smaller than the central values obtained in Ref.~\cite{Becirevic:2020rzi}. }
\label{fig:epssens}
\end{figure}
We have also observed that $R_0$, which is dominated by the ratio of decay widths, is more sensitive to NP than $R_1$. In consequence we put the focus on $R_0$ in the following.
 In Fig.~\ref{fig:epssens} we study the sensitivity of $R_0$ to NP signals of different chirality, $\e_X$, as a function of $q^2$. The $\e_X^{\rm{sens}}(q^2)$ are reported; they set the boundaries of the  regions of sensitivity. Red lines denote the Wilson coefficient values corresponding to a NP scale of 2 (dashed) and 10~TeV (solid). The plots reveal an excellent sensitivity to $\e_{L}$ and $\e_T$ (upper panels). For $\e_R$, the sensitivity is higher in the region of medium to high $q^2$ (lower left panel). We are also sensitive to quite small $\e_S$ (lower right panel). The only current to which this ratio is not sensitive is the pseudoscalar one. In this case the $\e_P$ signal is too small to be discerned from the SM, taking into account the $\mO(\e_P^2)$ theoretical error. 

Finally, from our analysis we conclude that the main obstacle to resolve a NP signal from the theoretical point of view is the LQCD error of the (axial)vector form factors, prevailing over the HQSS errors that accompany some NP contributions, and the SMEFT uncertainty, which can be sizeable for relatively large $\e_X$. 

\begin{figure}[h!]
    \centering
    \includegraphics[width=0.4\textwidth]{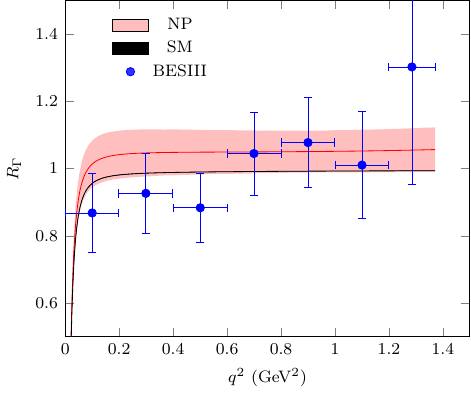}
    
    \caption{$\mu/e$ LFU ratio for the differential decay width ${\rm d}\Gamma/{\rm d}q^2$ as a function of $q^2$. The black band stands for the SM prediction from 1$\sigma$ errors in the LQCD form factors. The red band corresponds to the inclusion of NP with the empirical bounds obtained in Ref.~\cite{Becirevic:2020rzi}. The blue points are the experimental data from BESIII~\cite{BESIII:2023jxv}.}
    \label{fig:BES}
\end{figure}

One may also wonder about the experimental uncertainty and therefore the real sensitivity in practice. At present, the experimental error  is much larger than the theoretical one, at least in observables where SM uncertainties cancel. To illustrate this, we consider the integrated ratio $R$ (whose SM value is given in Eq.~\eqref{eq:RGamma}), setting all the $\e_{X\neq L}=0$ and accounting for the theoretical uncertainties. For instance, to be sensitive to $\e_L=1.2\times 10^{-2}$ as the central value extracted in  Ref.~\cite{Becirevic:2020rzi}, one would need the experimental error to be reduced to  10\% of its current size\footnote{Recall that we set all the $\e_{X\neq L}=0$ and account for the theoretical uncertainties.}.  
In addition, as discussed in Sec.~\ref{sec:pheno}, the tensions that arise in the description of $\Lambda_c\to\Lambda e^+ \nu_e$ with SM LQCD input compromise the extraction of $\e_X^\mu$ from $\Lambda_c\to\Lambda \mu^+ \nu_\mu$ data alone. 
On the other hand, recalling the aforementioned $\mO(m_\mu^2)$ suppression of the form factor uncertainty in the SM contribution to LFU ratios, we find that these ratios are more reliable in order to look for NP deviations, because systematic differences in the SM form factor largely cancel out. In Fig.~\ref{fig:BES} we report the LFU ratio for the differential decay width, defined in analogy with Eq.~(\ref{eq:Rdef}) as 
\begin{equation}\label{eq:RGamdef}
    R_{\Gamma}(\e^\mu_X)=\frac{d\Gamma/dq^2(\Lambda_c\to\Lambda \mu^+ \nu_\mu,\e^\mu_X)}{d\Gamma/dq^{2\,{\rm SM}}(\Lambda_c\to\Lambda e^+ \nu_e)} \ ,
\end{equation}
compared to the corresponding experimental result from BESIII~\cite{BESIII:2023jxv}. The theoretical result includes a black 1$\sigma$ SM band from LQCD and a red one accounting for the allowed range in the Wilson coefficients from Ref.~\cite{Becirevic:2020rzi}. In line with the previous comments about the ratio of integrated widths, the present experimental uncertainty is still insufficient to put competitive bounds on the $\e_\mu$ Wilson coefficients.   

\section{Sensitivity to new physics in the $\mu$ channel: $c^\mu_i$ functions}
\label{sec:Cs}

Here we report an analysis similar to the one performed for LFU ratios but at the level of the $c^\mu_i(q^2)$ functions. It is free of assumptions on the flavour hierarchy of Wilson coefficients. Furthermore, $c^\mu_i(q^2)$ would not receive any  contribution from the statistical experimental error in the $\Lambda_c\to\Lambda e^+ \nu_e$ decay. Unfortunately, the sensitivity is considerably reduced in comparison with the one expected for the LFU ratios explored in the previous section. We present only the cases where the NP perturbation was found to be relatively large in Sec.~\ref{sec:cix} (see Fig.~\ref{fig:ciX}). In consequence, we do not take into account (pseudo)scalar terms. 

\begin{figure}[h!]
\begin{subfigure}[t]{0.30\textwidth}
    \centering\includegraphics[width=\textwidth]{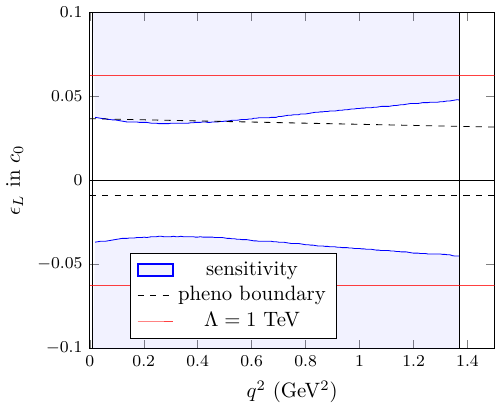}
    \caption{}
  \end{subfigure}\hfill
  \begin{subfigure}[t]{0.30\textwidth}
    \centering\includegraphics[width=\textwidth]{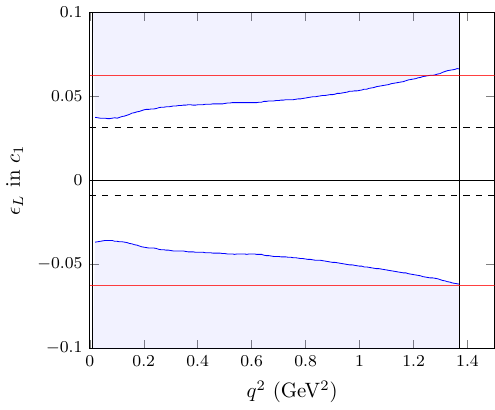}
   \caption{}
  \end{subfigure}\hfill
  \begin{subfigure}[t]{0.30\textwidth}
    \centering\includegraphics[width=\textwidth]{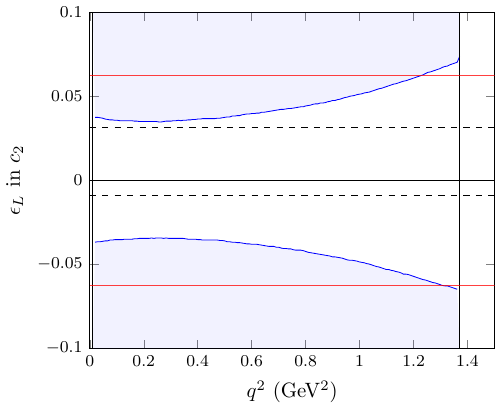}
    \caption{}
  \end{subfigure}\\
  \begin{subfigure}[t]{0.30\textwidth}
    \centering\includegraphics[width=\textwidth]{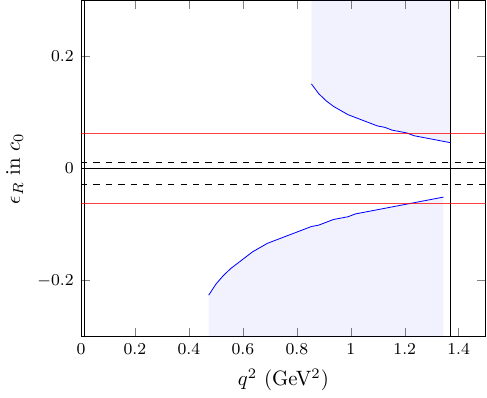}
    \caption{}
  \end{subfigure}\hfill
  \begin{subfigure}[t]{0.30\textwidth}
    \centering\includegraphics[width=\textwidth]{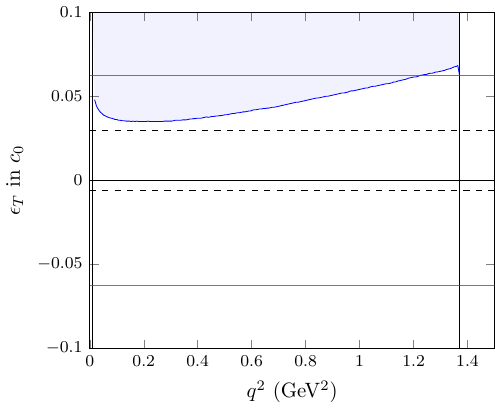}
   \caption{}
  \end{subfigure}\hfill
  \begin{subfigure}[t]{0.30\textwidth}
    \centering\includegraphics[width=\textwidth]{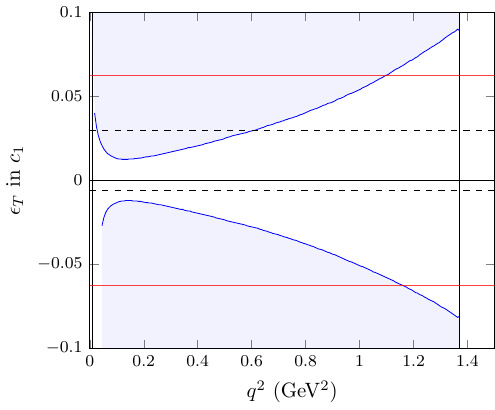}
   \caption{}
  \end{subfigure}
  \caption{The sensitivity regions to $\e_X$ in selected $c_i$ is represented by the light blue regions.
The solid blue line denotes $\e_X^{\rm{sens}}$ defined in Eq.~\eqref{eq:sens}. Panels (a), (b) and (c) show the sensitivity to $\e_L$ in $c_{0}$, $c_1$ and $c_2$, respectively. The (d) panel reports the poor sensitivity to the right-handed contribution in $c_0$ (note the different scale). The (e) and (f) panels display the sensitivity to $\e_T$ in $c_0$ and $c_1$ respectively. The dashed lines encompass the allowed region obtained in Ref.~\cite{Becirevic:2020rzi} from pseudoscalar meson decays. The red lines represent the values of Wilson coefficients for a $\Lambda=1$ TeV in the naive power counting. Black vertical lines delimit the phase space.}
\label{fig:sens}
\end{figure}

The left-handed contributions to $c_{0,1,2}$ are studied in Fig.~\ref{fig:sens} (a)-(c). In this case the truncation error is governed by $\e_L^2 c_i^{L\rm{l}}=2\e_L^2 c_i^{\rm{SM}}$ and the convergence is good, at least from the perspective of the uncertainty that we have defined. The error is dominated by the one of the LQCD vector form factors, even for high values of $\e_L$. The plots show that the study of $c_{0,1,2}$ for the present process is mostly  insensitive to the $\e_L$ range constrained by pseudoscalar meson decays in Ref.~\cite{Becirevic:2020rzi}.

The right-handed contributions are in general modest, being more prominent in $c_0$ than in $c_{1}$ and $c_2$. We show the sensitivity to $\e_R$ in $c_0$ in Fig.~\ref{fig:sens} (d). It is much poorer than in the other displayed cases. Indeed this observable is only sensitive to quite low
$\Lambda$ (below the TeV) in the naive power counting estimation\footnote{Approaching the electroweak scale, less NP models survive that have not been ruled out empirically.}. 
At low $q^2$, $c_0$ is not sensitive to $\e_R$ at all. In other words, the condition in Eq.~\eqref{eq:sens} is not satisfied. Actually, due to the small magnitude of $c_0^{R}$ 
at low $q^2$, $\e_R$ needs to be large to be distinguishable. However, at such  $\e_R$ values, the truncation error becomes large enough to make Eq.~\eqref{eq:sens} impossible to satisfy. The difference between positive and negative $\e_R^{\rm{sens}}$ is caused by the $\e_R^2 c_0^{R\rm{q}}$ term, which is important at high $\e_R$.  A similar scenario is encountered for the tensor contribution to $c_0$ studied in Fig.~\ref{fig:sens} (e). Now the decay under study is not sensitive to negative $\e_T$. This is due to the twofold impact of the quadratic term in Eq.~\eqref{eq:sens}, which is relatively large: not only does it notably increase $\Delta c_0(\e_T)$, it also conspires to cancel the linear contribution to $c_0(\e_T)$, having the opposite sign. As argued in the previous section, the quadratic contribution is not complete. Hence, the fact that $\e_{R,T}^2$ are large  implies that our sensitivity estimates are less solid in these cases. This is partially accounted by the truncation error. In all the other $c_i^{X}$ analysed in Fig.~\ref{fig:sens},  $\Delta c_i^{\rm{SMEFT}}$ is negligible for low values of $\e_X$ ($\e_X\leq 0.1$).

The most promising analysis turns out to be the study of $\e_T$ in $c_1$, Fig.~\ref{fig:sens} (f). As anticipated in Sec.~\ref{sec:cix}, the maximum in the tensor contribution coincides with a SM minimum (Fig.~\ref{fig:ciX} (b)). This leads to a good sensitivity in the low-medium $q^2$ region as can be seen in Fig.~\ref{fig:sens} (f). The quadratic coefficient $c_1^{T\rm{q}}$ is small and the truncation error is negligible (it is given by  
$\e_T^2 c_1^{T\rm{l}}$). The critical uncertainty in this scenario is actually the one of the LQCD vector and axial form factors $\Delta c_1^{\rm{LQCD}}$. In this case, the $\Lambda_c\to\Lambda \mu^+ \nu_\mu$ decay  is  sensitive to a part of the band allowed by pseudoscalar meson decays~\cite{Becirevic:2020rzi} without resorting to ratios. Finally, the tensor contribution to $c_2$ is not discussed because there is no linear term in $\e_T$.  

As a final remark, we stress that the key uncertainties to asses in order to gain sensitivity to NP with the $\Lambda_c$ semileptonic decay are the ones of the LQCD vector and axial form factors because they appear in the leading SM term. In the case of the tensor contribution to $c_0$, this error competes with the truncation one. In this sense, a more precise estimate of the SMEFT uncertainty, taking into account all terms at $\mO(1/\Lambda^4)$ would be very useful as well. In addition, it is clear that a reliable determination of the truncation error would provide more consistency to the analysis in general, including the left and right-handed contributions. In the tensor term the HQSS uncertainty is not so critical, since it is part of the NP correction (hence it is multiplied by $\e_T$) and is therefore small.

\section{\label{sec:conclusion}Conclusions and outlook}

We have studied the $\Lambda_c\to\Lambda \ell^{+} \nu_\ell$ decay, investigating the impact of NP currents in  the $\ell = \mu$ channel. To accomplish this task, an effective Hamiltonian for $c \to s \ell^+ \nu_\ell$, based on SMEFT has been employed. We have also relied on the LQCD results for the axial and vector $\Lambda_c\to\Lambda$ form factors obtained in Ref.~\cite{Meinel:2016dqj}. The form factors required to describe the $\Lambda_c\to\Lambda$ transitions driven by scalar, pseudoscalar and tensor operators have been expressed in terms of the SM axial and vector ones by taking advantage of the large $c$-quark mass. Differences arising from alternative approximations to implement this connection among form factors are treated as a theoretical uncertainty. 

The NP analyses are performed in terms of the $c_i(q^2)$ functions that drive the dependence on the angle between the $\Lambda$ and the $\mu^+$ in the leptonic rest frame. We have first obtained the $q^2$ shapes of the NP contributions to $c_i$, with their QCD uncertainty bands. 
Lepton flavour universality ratios received special attention. In the first place, owing to the LQCD form factor input, a SM ratio $\Gamma(\Lambda_c\to\Lambda \mu^+ \nu_\mu)/\Gamma(\Lambda_c\to\Lambda e^+ \nu_e)=0.96884(61)^{\rm LQCD}_{\rm{stat}}$ is obtained. LFU ratios are particularly advantageous in the search for NP due to the $\mO(m_\mu^2)$ suppression of the SM uncertainty. This is apparent in the reported $R_i = c_i^{\mu}/c_i^{e}$, $i=0,1,2$, ratios.

We have then studied the theoretical sensitivity to a potential NP current of a certain chirality in the $\mu$ channel, accounting also for the theoretical uncertainty in the SMEFT truncation. Assuming a  $\e^{\mu}\gg\e^{e}$ hierarchy, we have analysed different LFU ratios. In the ratio of integrated widths the sensitivity is already very high, probing the multi-TeV scale, except for the pseudoscalar interaction. However, to realise this theoretical expectation, the experimental uncertainty should be reduced as detailed in the work. 
In addition, we have shown that the ratios $R_0$ and $R_2$ 
have an excellent sensitivity to NP of any type except the pseudoscalar one. 
Finally, an analogous  sensitivity analysis has been performed on the $c_i^{\mu}$ observables themselves, which are free of the flavour hierarchy assumption. In this case the sensitivity varies notably depending on the NP chirality. In particular, the NP tensor contribution to $c_1$ displays a promising discovery potential.
   
We have identified the LQCD uncertainty of the axial and vector form factors as the most important one when it comes to improving  the sensitivity to NP. Besides, we have also seen that a better estimation of the SMEFT uncertainty would substantially improve the precision of this kind of analysis.

As a future development, the inclusion of right-handed neutrinos in the analysis would be interesting, since they are present in a large variety of NP models and can be implemented in a straightforward way within the framework established in Refs.~\cite{Penalva:2020xup} and \cite{Penalva:2021wye}.

\section*{Acknowledgements}
We thank Martín González Alonso for his valuable comments on the manuscript. This work has been partially supported by the Ministry of Science and Innovation of Spain under 
grants PID2020-112777GB-I00, PID2022-141910NB-I00 and PID2023-147458NB-C21, funded by MCIN/AEI/10.13039/501100011033, by the EU STRONG-2020 project under the program H2020-INFRAIA-2018-1, grant agreement no. 82409 and by Generalitat Valenciana's CIDEGENT/2019/015 and CIPROM/2023/59 programs. 

\appendix

\section{\label{sec:Eliecer}General framework and observables
}

With the aim of taking advantage of the general analytic results derived in  Refs.~\cite{Penalva:2020xup,Penalva:2021wye}, we rewrite the Hamiltonian of Eq.~(\ref{eq:H}) as 
\bea
H_{\rm eff}&=&\frac{4 G_F}{\sqrt2} V_{cs} \left[ C^V_{LR}{\cal O}^V_{LR}+
C^V_{RR}{\cal O}^V_{RR}+C^S_{LR}{\cal O}^S_{LR}+C^S_{RR}{\cal O}^S_{RR}
+C^T_{RR}{\cal O}^T_{RR} \right],
\label{eq:hnp}
\eea
where
\be
{\cal O}^V_{(L,R)R} = (\bar s \gamma_\alpha P_{L,R}\,c) 
(\bar \ell^c \gamma^\alpha P_R\nu^c_{\ell}), \, {\cal O}^S_{(L,R)R} = 
(\bar s\,  P_{L,R}\,c) (\bar \ell^c \, P_R\nu^c_{\ell}), \, {\cal O}^T_{RR} = 
(\bar s\, \sigma_{\alpha\beta} P_R\, c) (\bar \ell^c \sigma^{\alpha\beta} P_R\nu^c_{\ell})\,;
\label{eq:hnp2R}
\ee
$\ell^c,\nu_\ell^c$ are the charge-conjugated lepton fields. Taking into account that 
\bea
\bar\nu_\ell P_{R}\Gamma \ell\equiv-\bar\ell^c C\Gamma^TC^\dagger P_{R}
\nu^c_\ell,
\eea
and
\bea
C\Gamma^TC^\dagger=(-1)^n\Gamma\ ;\ 
\left\{\begin{array}{ll}n=0&,\,\Gamma=I\\
n=1&,\,\Gamma=\gamma_\alpha,\sigma_{\alpha\beta}\end{array}\right.,
\eea
where $C= i\gamma^2\gamma^0$ denotes the charge conjugation matrix, we identify
\be
\label{eq:wilsonconv}
C^V_{LR}=1+\epsilon^\ell_V,\ 
C^V_{RR}=\epsilon^\ell_R,\ 
C^S_{LR}=-\frac{\epsilon^\ell_S-\epsilon^\ell_P}2,\ 
C^S_{RR}=-\frac{\epsilon^\ell_S+\epsilon^\ell_P}2,\ 
C^T_{RR}=\epsilon^\ell_T \,.
\ee
Since the charge conjugated fields play for antiparticles the same role as the original fields for particles, all the expressions derived in Ref.~\cite{Penalva:2021wye} for $b \rightarrow c \ell^- \bar{\nu}_\ell$ are directly applicable in the understanding that, in the sum over the neutrino chiralities in Appendix D of that reference, only the part for $\chi=R$ has to be taken into account(see also Appendix E of that work).

In particular, one can write the differential decay width as
\bea
  \frac{{\rm d}^2\Gamma}{{\rm d}\omega {\rm d}s_{13}}& =& \frac{G^2_F|V_{cs}|^2 M m^{2}}{16\pi^3}\Big[{\cal A}(\omega)+{\cal B}(\omega) \frac{k\cdot p}{M^2}+{\cal C}(\omega) \frac{(k\cdot p)^2}{M^4}\Big]\,. \label{eq:Gamma}
\eea
$M$ and $m$ are the masses of the initial and final hadrons, respectively; $s_{13}\equiv (p-k)^2=M^2-2k\cdot p$; $\omega \equiv p\cdot p'/(M m)$ is the product of the hadron four-velocities, which is related to $q^2 = M^2 + m^{2} - 2 M m\omega$.
The functions ${\cal A}(\omega),\,{\cal B}(\omega)$ and ${\cal C}(\omega)$ are given in Appendix~D of Ref.~\cite{Penalva:2021wye} as a combination of Wilson coefficients\footnote{It is useful to bear in mind that from the definitions of Appendix A of Ref.~\cite{Penalva:2021wye} and Eq.~(\ref{eq:wilsonconv}) follows that 
$C^V_R=1+\epsilon^\ell_V+\epsilon^\ell_R,\ C^A_R=-(1+\epsilon^\ell_V-\epsilon^\ell_R),\ 
C^S_R=-\epsilon^\ell_S,\ 
C^P_R=-\epsilon^\ell_P,\ 
C^T_R=\epsilon^\ell_T$.}
and purely hadronic structure functions. The latter are given in terms of transition form factors: in the case of $1/2^+\to 1/2^+$ decays, as the one under study here, explicit expressions can be found in  Ref.~\cite{Penalva:2020xup}.

The $k\cdot p$ product can be written as
\be
k\cdot p=\frac
M2\Big(1-\frac{m^2_\ell}{q^2}\Big)(M_\omega+m\sqrt{\omega^2-1}\cos\theta)
\ee
with $M_\omega=M-m\omega$. Accordingly, 
\be
\label{eq:GammaCM}
 \frac{{\rm d}^2\Gamma}{{\rm d}\omega {\rm d}\cos\theta} = \frac{G^2_F|V_{c s}|^2 M^2
    m^{3}}{16\pi^3} \sqrt{\omega^2-1}
    \left(1-\frac{m_\ell^2}{q^2}\right)^2 
    [a_0(\omega) +a_1(\omega) \cos\theta + a_2(\omega)\cos^2\theta ]\,, 
\ee
where  $a_{0,1,2}(\omega)$  are linear combinations of ${\cal A}(\omega),\,{\cal B}(\omega)$ and ${\cal C}(\omega)$ (see 
Eq.~(18) of Ref.~\cite{Penalva:2020xup}). Finally, the $c_i$ functions introduced in Eq.~(\ref{eq:ci}) correspond to $a_i$ up to a normalization: 
\begin{equation}\label{eq:ci-ai}
    c_i=\frac{G^2_F M
    m^{2}}{32\pi^3}
  \sqrt{\omega^2-1}\left(1-\frac{m_\ell^2}{q^2}\right)^2 \frac{a_i}{\Gamma_{\rm{tot}}}.
\end{equation}
\par 
The $c_i$ functions can be extracted from measurable quantities:
the differential width $d\Gamma/dq^2$, the forward-backward asymmetry $\mA_{\rm{FB}}$, and the convexity $\mA_{\pi/3}$.  They are related as follows~\cite{Becirevic:2020rzi,Colangelo:2021dnv,Penalva:2020xup}, 
\begin{equation}\label{eq:diffwidth}
    \frac{1}{\Gamma_{\rm tot}}\frac{{\rm d}\Gamma}{{\rm d}q^2}=\abs{V_{cs}}^2\,2(c_0+c_2/3) \ , 
\end{equation}
\bea\label{eq:fb}
    \mA_{\rm{FB}}&\equiv & \left(\frac{{\rm d}\Gamma}{{\rm d}q^2}\right)^{-1} \left[\int_0^1 {\rm d} \cost  \frac{{\rm d}^2\Gamma}{{\rm d}q^2 {\rm d} \cost} -  \int_{-1}^0 {\rm d} \cost  \frac{{\rm d}^2\Gamma}{{\rm d}q^2 {\rm d} \cost} \right]\no\\
    &=&\frac{c_1}{2(c_0+ c_2/3)}\no\\
&=&c_1 \,\abs{V_{cs}}^2  \left(\frac{{\rm d}\mB}{{\rm d}q^2}\right)^{-1} \ ,
\eea
\bea\label{eq:convexity}
    \mA_{\pi/3}&\equiv & \left(\frac{{\rm d}\Gamma}{{\rm d}q^2}\right)^{-1} \Bigg[\int_{1/2}^1 {\rm d} \cost  \frac{{\rm d}^2\Gamma}{{\rm d}q^2 {\rm d} \cost} -  \int_{-1/2}^{1/2} {\rm d} \cost  \frac{{\rm d}^2\Gamma}{{\rm d}q^2 {\rm d} \cost}+ \int_{-1}^{-1/2} {\rm d} \cost  \frac{{\rm d}^2\Gamma}{{\rm d}q^2 {\rm d} \cost}  \Bigg]\no\\
    &=&\frac{c_2}{4(c_0+ c_2/3)}\no\\
    &=& \frac{1}{2}\, c_2 \,\abs{V_{cs}}^2  \left(\frac{{\rm d}\mB}{{\rm d}q^2}\right)^{-1} \ .
\eea

\section{\label{sec:HQSS} Relations among form factors}

Here we describe the approximations and prescriptions adopted to express scalar, pseudoscalar and tensor $\Lambda_c \to \Lambda$ transition form factors in terms of vector and axial ones, which have been determined using LQCD~\cite{Meinel:2016dqj}. All of them rely on the fact that  $m_c/\Lambda_{\rm QCD}>>1$.

\subsection{OSHQ approximation}

Let us consider the transition between a heavy ($H$) and a light ($L$) spin $1/2$ baryons induced by vector current $J^{\mu}_V(x)=\bar{l}(x)\gamma^{\mu}h(x)$, in terms of heavy and light quark fields, $h$ and $l$. We assume that 
\begin{itemize}
    \item $h$ can be treated as a free field obeying the Dirac equation $(i\slashed{\partial}-m_{h})h(x)=0$, with $m_h$ the heavy quark mass.
    \item $\ket{H(s,p)}$ can be factorized as $\ket{H(s,p)}=\ket{h(\tilde{s},\tilde{p})}\ket{\lambda}$, where $\ket{\lambda}$ is a residual light system. $\tilde{p}$ stands for the fraction of the $H$ momentum $p$ carried by the heavy quark, satisfying the on-shell condition $\tilde{p}^2=m_h^2$; $\tilde{s}$ is the corresponding fraction of the spin projection $s$ of $H$. 
\end{itemize}    
    
Relying on the standard second quantization decomposition of $h(x)$, these two premises lead to
    \begin{equation}
        \bra{L(s',p')}J_V^\mu(x)\ket{H(s,p)}= \bra{L} \bar{l}(x)\gamma^\mu u(\tilde{s},\tilde{p}) e^{-i\tilde{p}\cdot x}  \ket{0}_h \ket{\lambda} 
        \label{eq:MatElV}
    \end{equation}
for the matrix element of the vector current. Proceeding analogously for the scalar current $J_S=\bar{l}(x)h(x)$ one gets
\begin{equation}
\bra{L}J_S(x)\ket{H}=  \bra{L} \bar{l}(x) u(\tilde{s},\tilde{p}) e^{-i\tilde{p}\cdot x}  \ket{0}_h \ket{\lambda}   
\label{eq:MatElA}
\end{equation}

We further assume that 
    \begin{itemize}
    \item $\tilde{p}\approx p$ or, more specifically, that the heavy quark spinor, $u(\tilde{s},\tilde{p})$, satisfies $\slashed{p}u(\tilde{s},\tilde{p})=m_h u(\tilde{s},\tilde{p})$. 
    \end{itemize}
    
We can therefore contract
    \bea\label{eq:pmu}
        p_\mu \bra{L}J_V^\mu(x)\ket{H}= \bra{L} \bar{l}(x)\slashed{p} u(\tilde{s},\tilde{p}) e^{-i\tilde{p}\cdot x}  \ket{0}_h \ket{\lambda} =m_h \bra{L} \bar{l}(x) u(\tilde{s},\tilde{p}) e^{-i\tilde{p}\cdot x}  \ket{0}_h \ket{\lambda} \ .
    \eea
This result, together with Eqs.~(\ref{eq:MatElV},\ref{eq:MatElA}), provides the following connection between the matrix elements of the vector and scalar currents
\begin{equation}
    p_\mu \bra{L}J_V^\mu(x)\ket{H}=m_h  \bra{L}J_S(x)\ket{H} \,.
\end{equation}
Then, identifying $H=\Lambda_c$, $L=\Lambda$, $h=c$, $l=s$ one finds that
\begin{equation}
    F_S \bar{u}_\Lambda(p')u_{\Lambda_c}(p)=\frac{1}{m_c}\bar{u}_\Lambda(p')p_\mu \Gamma^\mu_V  u_{\Lambda_c(p)} \,,
    \label{eq:FS}
\end{equation}
where 
\begin{equation}
    \Gamma^\mu_V=\gamma^\mu F_1+\frac{p^\mu}{M}F_2+\frac{p^{\prime\mu}}{m}F_3 \,,
\end{equation}
[see Eq.~(\ref{eq:FactoresForma})]. With the help of the Dirac equations for baryons, Eq.~(\ref{eq:FS}) allows to write $F_S$ in terms of $F_{1-3}$. This relation is given in Eq.~\eqref{eq:ffLuis} in the $m_c/M \to 1$ limit. 

Replacing $J_V(x)$ by the axial current $J^{\mu}_A(x)=\bar{l}(x)\gamma^{\mu}\gamma_5 h(x)$ and $J_S(x)$ by the pseudoscalar one $J_P=\bar{l}(x) \gamma_5 h(x)$ one obtains 
\begin{equation}
    F_P \bar{u}_\Lambda(p')\gamma_5 u_{\Lambda_c}(p)= -\frac{1}{m_c}\bar{u}_\Lambda(p')p_\mu \Gamma^\mu_A  u_{\Lambda_c(p)} \,,
    \label{eq:FP}
\end{equation}
with
\begin{equation}
    \Gamma^\mu_A=\left( \gamma^\mu G_1+\frac{p^\mu}{M}G_2+\frac{p^{\prime\mu}}{m}G_3 \right)\gamma_5  \,.
\end{equation}
The resulting expression for $F_P$ in terms of axial form factors $G_{1-3}$ is also given in Eq.~\eqref{eq:ffLuis}. 

Next we repeat the procedure with tensor currents $J_T^{\mu\nu}(x) = \bar{l}(x)\sigma^{\mu\nu}h(x)$ and $J_{TA}^{\mu\nu}(x) = \bar{l}(x)\sigma^{\mu\nu}\gamma_5 h(x)$ and obtain
\begin{eqnarray}
    \bar{u}_\Lambda(p') \left[ (p_\alpha \Gamma_V^\alpha)p^\mu - m_c^2 \Gamma_V^\mu \right] u_{\Lambda_c}(p) &=& i m_c \bar{u}_\Lambda(p') \Gamma^{\mu\nu}p_\nu u_{\Lambda_c}(p) \label{eq:Tensor1}
    \\[0.2cm]
     \bar{u}_\Lambda(p') \left[ (p_\alpha \Gamma_A^\alpha)p^\mu - m_c^2 \Gamma_A^\mu \right] u_{\Lambda_c}(p) &=& - i m_c \bar{u}_\Lambda(p') \Gamma_A^{\mu\nu}p_\nu u_{\Lambda_c}(p) \label{eq:Tensor2}
\end{eqnarray}
where ($\e_{0123}=+1$)
\begin{equation}
    \Gamma_A^{\mu\nu} \equiv -\frac{i}{2} \epsilon^{\mu\nu\alpha\beta} \Gamma_{\alpha\beta} \ ,
\end{equation}
thanks to the relation
\begin{equation}
\sigma^{\mu\nu}\gamma_5=-\frac{i}{2}\e^{\mu\nu\alpha\beta}\sigma_{\alpha\beta} \ ,
\end{equation}
and 
\begin{equation}
    \Gamma^{\alpha\beta} = \frac{i}{M^2}
 \left(p^\alpha p^{\prime \beta}-p^\beta p^{\prime \alpha}\right) T_1
 + \frac{i}{M}\left(\gamma^\alpha p^\beta
  -\gamma^\beta p^\alpha  \right) T_2 
 + \frac{i}{M}\left(\gamma^\alpha p^{\prime \beta}
-\gamma^\beta p^{\prime \alpha}\right) T_3
+ \sigma^{\alpha\beta} T_4 \,.
\end{equation}
In the $m_c/M \to 1$ limit, Eqs.~(\ref{eq:Tensor1},\ref{eq:Tensor2}) lead to a consistent and independent set of four equations for $T_{1-4}$ in terms of $F_{1,3}$ and $G_{1,3}$, completing the OSHQ prescription of Eq.~\eqref{eq:ffLuis}.
\subsection{HQSS relations}

Following Sec.~2.10 of Ref.~\cite{Manohar:2000dt}, at leading order in $1/m_c$, one can write that 
\begin{equation}
    \langle\Lambda (p')| \bar s \gamma c | \Lambda_c (p)\rangle \to \langle\Lambda (p')| \bar s \gamma c_v | \Lambda_c (v)\rangle = \bar{u}_{\Lambda}(p') \Gamma u_{\Lambda_c}(v) \,
    \label{eq:HQmatrixelm1}
\end{equation}
where $\slashed{v} u_{\Lambda_c}(v) = u_{\Lambda_c}(v)$ and $v=p/M$; $\gamma = \mathds{1},\,\gamma_5,\,\gamma^\mu,\,\gamma^\mu\gamma_5,\,\sigma^{\mu\nu}$ while $\Gamma$ denotes the corresponding representation in terms of form factors given by Eq.~(\ref{eq:FactoresForma}). On the other hand, in general,
\begin{equation}
    \langle\Lambda (p')| \bar s \gamma c_v | \Lambda_c (v)\rangle = \bar{u}_{\Lambda}(p') (\tilde{F}_1 + \tilde{F}_2\slashed{v}) \gamma u_{\Lambda_c}(v) \,.
\end{equation}
Therefore, in the HQSS limit, all form factors introduced in Eq.~(\ref{eq:FactoresForma}) can be expressed in terms of two, $\tilde{F}_{1,2}$. In fact, straightforward Dirac algebra manipulations lead to 
\begin{description}
    \item[$\gamma = \mathds{1}$] $F_S = \tilde{F}_1 + \tilde{F}_2$ 
    \item[$\gamma = \gamma_5$] $F_P = \tilde{F}_1 - \tilde{F}_2$
    \item[$\gamma = \gamma^\mu$] $F_1 = \tilde{F}_1 - \tilde{F}_2$, $F_2 = 2 \tilde{F}_2$, $F_3 = 0$
    \item[$\gamma = \gamma^\mu \gamma_5$] $G_1 = \tilde{F}_1 + \tilde{F}_2$, $G_2 =  2 \tilde{F}_2$, $G_3 = 0$
    \item[$\gamma = \sigma^{\mu\nu}$] $T_1 = T_3 = 0$, $T_2 = - 2 \tilde{F}_2$, $T_4 = \tilde{F}_1 + \tilde{F}_2$.
\end{description}
These equations allow to write scalar, pseudoscalar and tensor form factors as a function of $F_{1,2}$ or $G_{1,2}$, giving rise to the models denoted by HQSSV [Eq.~(\ref{eq:ffV})] or HQSSA [Eq.~(\ref{eq:ffA})], respectively.

\bibliography{main}

\end{document}